\documentclass[onecolumn,useAMS,usenatbib,twocolumn]{mn2e}

\usepackage{graphicx}
\usepackage{subfigure}
\usepackage{xcolor}
\usepackage{mathtext,bbm,amsmath,amsfonts,amssymb,indentfirst,syntonly,graphicx}
\usepackage{mathtools}
\usepackage{slashbox}
\usepackage[english]{babel}
\usepackage{calc}
\usepackage{tikz}

\def\bc{\begin{center}}
\def\ec{\end{center}}
\def\be{\begin{eqnarray}}
\def\ee{\end{eqnarray}}

\title[Testing isotropy of the Universe using JLA]{Testing the isotropy of the Universe by using the JLA compilation of type-Ia supernovae}
\author[H.-N. Lin, S. Wang, Z. Chang and X. Li]
        {Hai-Nan Lin$^{1}$\thanks{e-mail: linhainanjyzjcn@163.com.},
        Sai Wang$^{2}$\thanks{e-mail: wangsai@itp.ac.cn.},
        Zhe Chang$^{3}$\thanks{e-mail: changz@ihep.ac.cn.} and
        Xin Li$^{1,2}$\thanks{e-mail: lixin1981@cqu.edu.cn.}\\
$^{1}$Department of Physics, Chongqing University, Chongqing 401331, China\\
$^{2}$State Key Laboratory of Theoretical Physics, Institute of Theoretical Physics, Chinese Academy of Sciences, Beijing 100190, China\\
$^{3}$Institute of High Energy Physics, Chinese Academy of Sciences, Beijing 100049, China}
\begin{document}

\date{Accepted xxxx; Received xxxx; in original form xxxx}

\pagerange{\pageref{firstpage}--\pageref{lastpage}} \pubyear{2015}

\maketitle

\label{firstpage}

\begin{abstract}
We probe the possible anisotropy of the Universe by using the JLA compilation of type-Ia supernovae. We apply the Markov Chain Monte Carlo (MCMC) method to constrain the amplitude and direction of anisotropy in three cosmological models. For the dipole-modulated $\Lambda$CDM model, the anisotropic amplitude is consistent with zero at $68\%$ C.L., and has an upper bound $A_D<1.98\times10^{-3}$ at $95\%$ C.L. Regardless of much larger uncertainty, we find the dipole direction of JLA is amazingly opposite to that of Union2. Similar results are found for the dipole-modulated $w$CDM and CPL models. Thus, the Universe is still well consistent with the isotropy according to the JLA compilation.
\end{abstract}

\begin{keywords}
supernovae: general \--- large-scale structure of Universe
\end{keywords}

\section{Introduction}\label{sec:introduction}

The cosmological principle, which is one of the foundations in modern cosmology, says that the Universe is homogeneous and isotropic at large enough scales \citep{Dodelson:2003ft}. It is well consistent with the present observational data, such as the cosmic microwave background (CMB) radiation from Wilkinson Microwave Anisotropy Probe (WMAP) \citep{Bennett:2012zja,Hinshaw:2012aka} and Planck satellite \citep{Ade:2015xua,Ade:2015lrj}. Until now, the cosmological observations are still in accordance with the cosmological constant plus cold dark matter ($\Lambda$CDM) model which is based on the cosmological principle. Thus, the $\Lambda$CDM model becomes the leading model in modern cosmology.

Despite the great successes it achieved, the $\Lambda$CDM model still faces certain challenges \citep{Perivolaropoulos:2008ud,Perivolaropoulos:2011hp,Mariano:2012ia}. As the improvements of accuracy, it is found from a large amount of observations that the Universe might deviate from statistical isotropy. These include the alignment of low multipoles in the angular power spectrum of CMB temperature fluctuations \citep{Tegmark:2003ve,Bielewicz:2004en,Copi:2010na,Frommert:2009qw}, the hemispherical power asymmetry of CMB temperature anisotropy \citep{Bennett:2012zja,Ade:2013nlj,Eriksen:2003db,Hansen:2004vq,Akrami:2014eta,Quartin:2014yaa}, the spatial variation of the electromagnetic fine-structure constant \citep{Webb:2010hc,King:2012id,Molaro:2013saa}, the large-scale alignment of the quasar polarization vectors \citep{Hutsemekers:2000fv,Hutsemekers:2005iz}, the large-scale bulk flow beyond the prediction of $\Lambda$CDM model \citep{Kashlinsky:2008ut,Kashlinsky:2009dw,Watkins:2008hf}, and so on. All of these phenomena arouse us to rethink the validity of the cosmological principle. If the cosmological principle is proven to be failed, the modern cosmology should be rewritten.

Due to their consistent absolute magnitudes, type-Ia supernovae (SNe Ia) are regarded as the ideal distance indicators to trace the accelerated expansion of the Universe. In fact, they have been widely used to search for the anisotropic signals in the Universe. Especially, a statistic based on the extreme value theory shows that the gold data set is consistent with the isotropy \citep{Gupta:2007pb}. The study \citep{Blomqvist:2010ky} on the angular covariance function of supernova magnitude fluctuations is consistent with zero dark energy fluctuations by using the Union2 compilation \citep{Amanullah:2010vv}. A ``\,residual'' statistic shows that the isotropic $\Lambda$CDM model is not consistent with the Union2 data with $z<0.05$ at $2-3\sigma$ \citep{Colin:2010ds}. There are no significant evidence for deviations from the isotropy in the anisotropic Bianchi-I cosmology \citep{Campanelli:2010zx,Schucker:2014wca}, Bianchi-III and Kantowski-Sachs metrics \citep{Koivisto:2010dr}, and Randers-Finsler cosmology \citep{Chang:2014wpa,Chang:2013xwa,Chang:2013zwa}. The hemisphere comparison is used to study the Union2 data and shows certain preferred directions \citep{Schwarz:2007wf,Antoniou:2010gw,Cai:2011xs,Kalus:2012zu,Yang:2013gea,Chang:2014nca}. By dividing the Union2 supernovae into 12 subsets according to their galactic coordinates, a dipole of the deceleration parameter is preferred at more than $2\sigma$ level \citep{Zhao:2013yaa}. By combining the data of Union2 and gamma-ray bursts, the isotropic $\Lambda$CDM model is well permitted \citep{Cai:2013lja} while the anisotropic Finsler cosmology is preferred at around $2\sigma$ \citep{Chang:2014jza}. By using the data of Union2.1 \citep{Suzuki:2011hu} and gamma-ray bursts, a model-independent way shows a dipolar anisotropy at more than $2\sigma$ \citep{Wang:2014vqa}. It has been found that there may be certain correlation between the fine structure dipole and the dark energy dipole \citep{Mariano:2012wx,Li:2015uda}. A fully-Bayesian method was developed to remove the systematics in the Union datasets and the anisotropic cosmology does not seem to be reflected \citep{Heneka:2013hka}.

Recently, a new sample of SNe Ia was released by the SDSS collaboration, which is called the ``\,joint light-curve analysis'' (JLA) compilation \citep{Betoule:2014frx}. Compared to previous compilations such as Union2 \citep{Amanullah:2010vv} and Union2.1 \citep{Suzuki:2011hu}, the number of SNe Ia in the JLA compilation is highly enlarged and the systematic uncertainties are significantly reduced. Recently, the JLA SNe Ia have been used to probe the anisotropic Hubble diagram in Bianchi type I cosmology \citep{Schucker:2014wca} and test the cosmological principle \citep{Bengaly:2015dza}. However, the work \citep{Bengaly:2015dza} did not consider the full covariance matrix between SNe Ia. In this paper, we use the JLA compilation to restudy the anisotropic Hubble diagram of the Universe. Unlike certain previous works which have neglected the correlations between any two SNe Ia, we make use of the full covariance matrix to construct the likelihood (or chi-square). In addition, we use the method of MCMC sampling in our analysis. It has been shown that the statistical significance of the previously claimed evidence for a preferred direction could be highly lowered if the full covariance matrix of SNe Ia is considered in the Union2 compilation \citep{Jimenez:2014jma}. We want to see whether the anisotropic signals in the accelerated expansion of the Universe still exist in the newly released JLA compilation.

The rest of the paper is arranged as follows. In section \ref{sec:modelsandmethodology}, we briefly introduce the JLA dataset, and present the anisotropic cosmological models and the numerical method used in our analysis. In section \ref{sec:results}, we give constraints on the anisotropic amplitudes and directions for the anisotropic expansion of the Universe. Finally, our conclusions are given in section \ref{sec:conclusion}.

\section{Data and Methodology}\label{sec:modelsandmethodology}

The anisotropic expansion of the Universe can be induced by assuming that the dark energy has anisotropic repulsive force \citep{ArmendarizPicon:2004pm,Koivisto:2008xf,Salehi:2015ira}, or the background spacetime has a certain preferred direction \citep{Chang:2013xwa,Chang:2013zwa,Li:2013vea,Li:2015uda,Schucker:2014wca}, and so on. In this paper, we assume a dipole modulation to describe the Universe deviating from the isotropic background. Phenomenologically, the direction-dependent distance modulus can be given as
\begin{equation}
\label{muth}
\mu_{\rm th}=\bar{\mu}_{\rm th}\left(1+A_D (\hat{\textbf{n}}\cdot\hat{\textbf{p}})\right),
\end{equation}
where $A_D$ denotes the amplitude of the dipole modulation, $\hat{\textbf{n}}$ is the dipole direction, $\hat{\textbf{p}}$ is the unit 3-vector pointing towards the supernova, and $\bar{\mu}_{\rm th}$ denotes the theoretical distance modulus predicted by the isotropic $\Lambda$CDM, $w$CDM or CPL models. Here the anisotropic amplitude $A_D$ is assumed to be a constant over the whole redshift range. In the galactic coordinates, the dipole direction $\hat{\textbf{n}}$ can be parameterized as $(l,b)$, where $l$ and $b$ are the longitude and latitude, respectively. In such a parametrization, we have $\hat{\textbf{n}}=\cos(b)\cos(l)\hat{\textbf{i}}+\cos(b)\sin(l)\hat{\textbf{j}}+\sin(b)\hat{\textbf{k}}$, where $\hat{\textbf{i}}$, $\hat{\textbf{j}}$, $\hat{\textbf{k}}$ are the unit vectors along the axes of a Cartesian coordinates system. The position of the $i$th supernova with galactic coordinates $(l_i,b_i)$ can be written as
$\hat{\textbf{p}}_i=\cos(b_i)\cos(l_i)\hat{\textbf{i}}+\cos(b_i)\sin(l_i)\hat{\textbf{j}}+\sin(b_i)\hat{\textbf{k}}$.

In the spatially-flat isotropic background spacetime, we can express the luminosity distance $d_L(z)$ of a supernova in terms of the redshift $z$,
\begin{equation}\label{luminositydistance}
d_L(z)=\frac{1+z}{H_0}\int_0^{z} \frac{dz^\prime}{E(z^\prime)}\ ,
\end{equation}
where $H_0=100h~\rm{km}~\rm{s}^{-1}~\rm{Mpc}^{-1}$ is the Hubble constant, and $E(z)$ is a function of redshift. The isotropic distance modulus $\bar{\mu}_{\rm th}$ can be given by
\begin{equation}
\bar{\mu}_{\rm th}=5\log_{10}\frac{d_L}{10~\rm{pc}}.
\end{equation}
In equation (\ref{luminositydistance}), the quantity $E(z)$ is a function of redshift $z$. The expression of $E(z)$ depends on a specific cosmological model. In the $\Lambda$CDM model, it can be expressed as
\begin{equation}
E^2(z)=\Omega_{m}(1+z)^3+(1-\Omega_{m})\ ,
\end{equation}
where $\Omega_{m}$ is the energy density of matter today. In the $w$CDM model, it can be expressed as
\begin{equation}
E^2(z)=\Omega_{m}(1+z)^3+(1-\Omega_{m})(1+z)^{3(1+w)}\ ,
\end{equation}
where $w\equiv p/\rho$ denotes the equation of state of dark energy. In the Chevallier-Polarski-Linder (CPL) parametrization \citep{Chevallier:2000qy,Linder:2002et}, the equation of state of dark energy is redshift-dependent, and it is parameterized by $w=w_0+w_1z/(1+z)$. In this case, $E(z)$ can be expressed as
\begin{equation}
E^2(z)=\Omega_{m}(1+z)^3+(1-\Omega_{m})(1+z)^{3(1+w_0+w_1)}\exp{\left(-3w_1\frac{z}{1+z}\right)}\ .
\end{equation}

In this paper, we use the most recently published JLA compilation \citep{Betoule:2014frx} of SNe Ia to constrain the anisotropy of the Universe. The JLA compilation consists of 740 well-calibrated SNe Ia in the redshift range of $z\in[0.01,1.30]$. It is a collection of several low-redshift samples, all three seasons from the SDSS-II, three years from SNLS, and a few high-redshift samples from the Hubble Space Telescope (HST). All of the SNe Ia have high-quality light curves, so their distance moduli can be abstracted with high precision. The positions of SNe Ia in the sky of equatorial coordinates system can be found at the website of IAU Central Bureau for Astronomical Telegrams\footnote{http://www.cbat.eps.harvard.edu/lists/Supernovae.html}.
To compare with others' work, we transform the positions of SNe Ia into the galactic coordinates.

From the observational point of view, the distance modulus of a SN Ia can be abstracted from its light curve through the empirical linear relation \citep{Betoule:2014frx}
\begin{equation}\label{muobs}
  \hat{\mu}=m_B^{*}-(M_B-\alpha\times X_1+\beta \times \mathcal{C}),
\end{equation}
where $m_B^*$ is the observed peak magnitude in rest frame $B$ band, $M_B$ is the absolute magnitude depending on the host galaxy properties complexly, $X_1$ is the time stretching of the light curve, and $\mathcal{C}$ is the supernova color at maximum brightness. The three light-curve parameters $m_B^*$, $X_1$ and $\mathcal{C}$ are different from one supernova to other one and can be derived directly from the light curves. The two nuisance parameters $\alpha$ and $\beta$ are assumed to be constants for all the supernovae.

For the JLA samples, the isotropic luminosity distance of a supernova can be given as
\begin{equation}
d_L(z_{\rm hel},z_{\rm cmb})=\frac{1+z_{\rm hel}}{H_0}\int_0^{z_{\rm cmb}} \frac{dz'}{E(z')},
\end{equation}
where $z_{\rm cmb}$ and $z_{\rm hel}$ denote the CMB frame redshift and heliocentric redshift, respectively. Then we can obtain the anisotropic distance modulus $\mu_{\rm th}$ in equation (\ref{muth}). Using the observed distance modulus $\hat{\mu}$ in equation (\ref{muobs}), the anisotropic cosmological models can be fitted to the JLA dataset by using the chi-square as
\begin{equation}\label{chijla}
\chi^2_{\rm JLA}=\left(\hat{\mu}-\mu_{\rm th}\right)^{\dagger}C^{-1}\left(\hat{\mu}-\mu_{\rm th}\right),
\end{equation}
where $C$ is the covariance matrix of $\hat{\mu}$, and it is presented in \citet{Betoule:2014frx}.

In order to directly compare with previous works, we also apply our method to the Union2 \citep{Amanullah:2010vv} dataset. The Union2 data set consists of 557 SNe Ia with well-observed redshift in the range of $z \in [0.015, 1.4]$. The distance moduli and their uncertainties are extracted from the SALT2 light-curve fitter. The directions of SNe Ia are well localized in the sky of the equatorial coordinates \citep{Blomqvist:2010ky}. The data in Union2 are usually assumed to be uncorrelated. Thus, the chi-square of Union2 is simplified to
\begin{equation}\label{union2}
\chi^2_{\rm Union2}=\sum\left(\frac{\hat{\mu}-\mu_{\rm th}}{\sigma_{\mu}}\right)^2,
\end{equation}
where the summation runs over all the SNe Ia data.

In this paper, we employ the Markov Chain Monte Carlo (MCMC) method to estimate the model parameters. The joint likelihood is given by $\mathcal{L}\propto\exp(-\chi^2/2)$. The nuisance parameters  such as $\alpha$ and $\beta$ are marginalized. We modify the publicly available Cosmological Monte Carlo sampler (CosmoMC) \citep{Lewis:2002ah} to estimate the background parameters and anisotropic parameters. For $\Lambda$CDM model, the isotropic parameter can be well constrained by supernovae data. For $w$CDM and CPL models, however, the isotropic parameters can't be well constrained by using the supernovae data only. Following the method in \citet{Betoule:2014frx}, therefore, we combine the supernovae data with the Planck~2013 results of CMB temperature anisotropy (Planck2013), the WMAP9 observations of CMB polarizations (WP), and the SDSS-III BOSS DR11 Baryon acoustic oscillations data (BAO) to constrain the isotropic parameters. Once the isotropic parameters are given, we can fix them and fit the anisotropic parameters with JLA SNe Ia only.

\section{Results}\label{sec:results}

\begin{figure}
  \centering
    \subfigure[]{
    \label{fig:subfig:a1} 
    \includegraphics[width=1.6in]{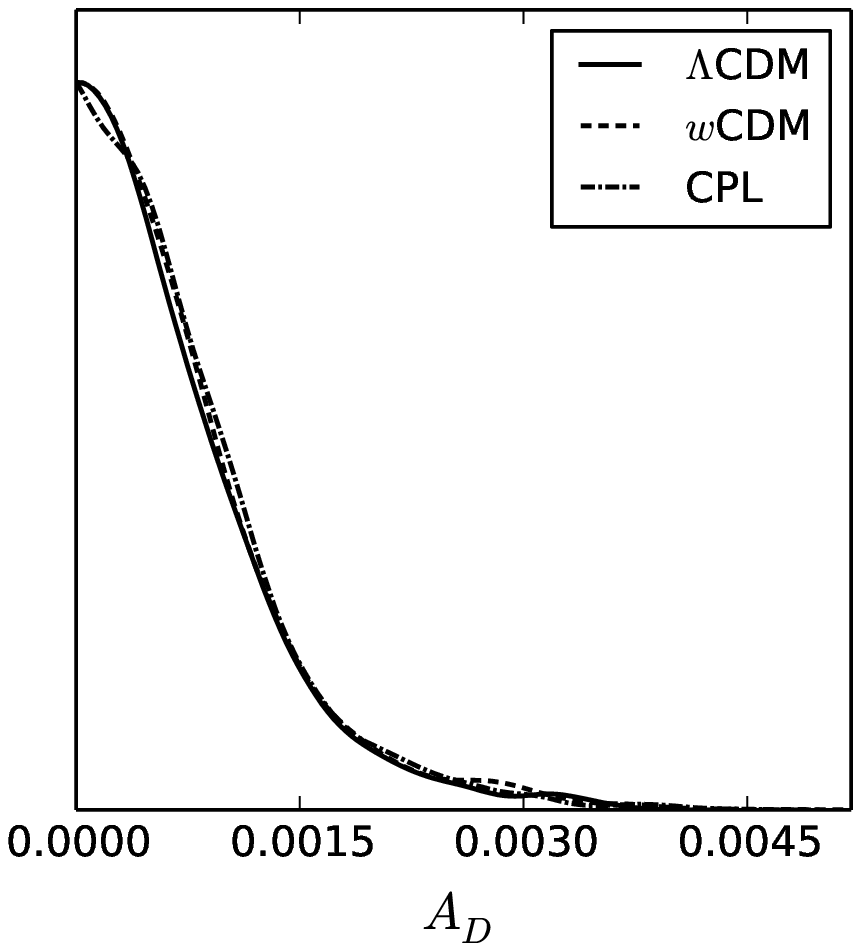}}
  \subfigure[]{
    \label{fig:subfig:b1} 
    \includegraphics[width=1.56in]{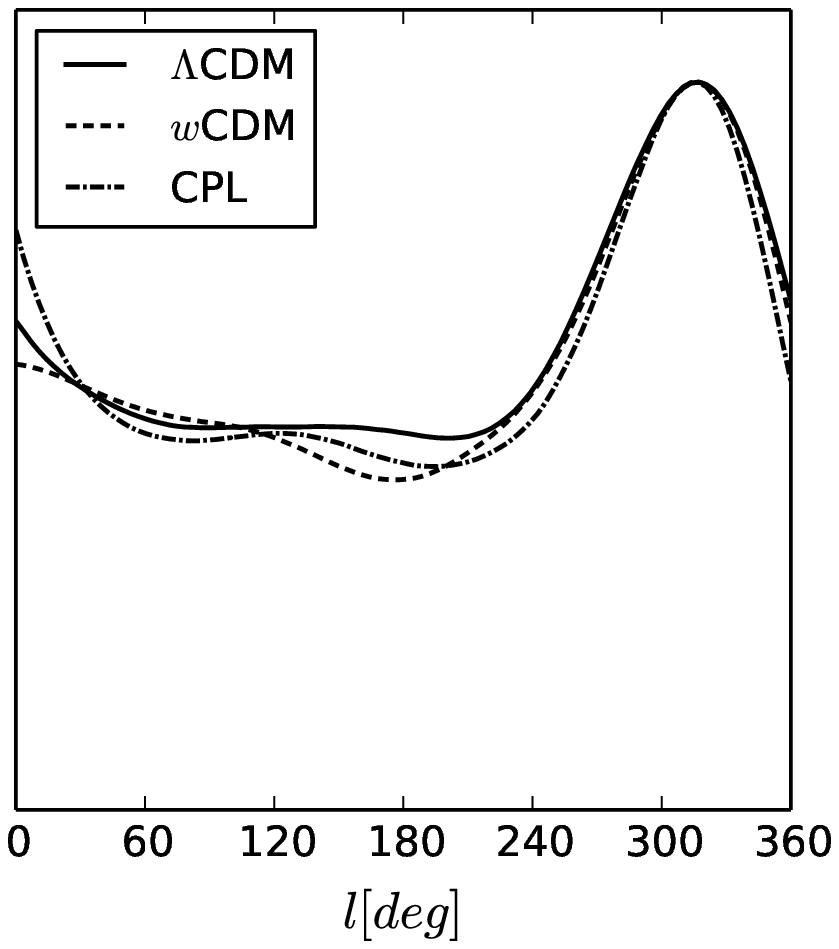}}
  \subfigure[]{
    \label{fig:subfig:c1} 
    \includegraphics[width=1.6in]{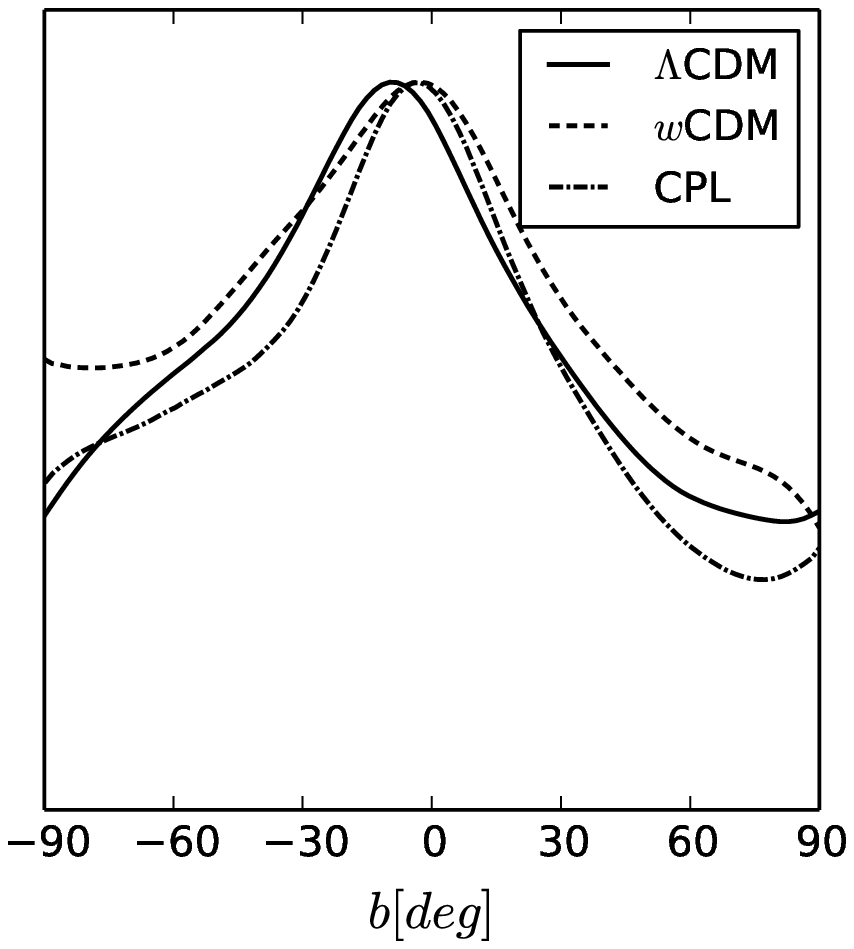}}
  \subfigure[]{
    \label{fig:subfig:c1} 
    \includegraphics[width=1.5in]{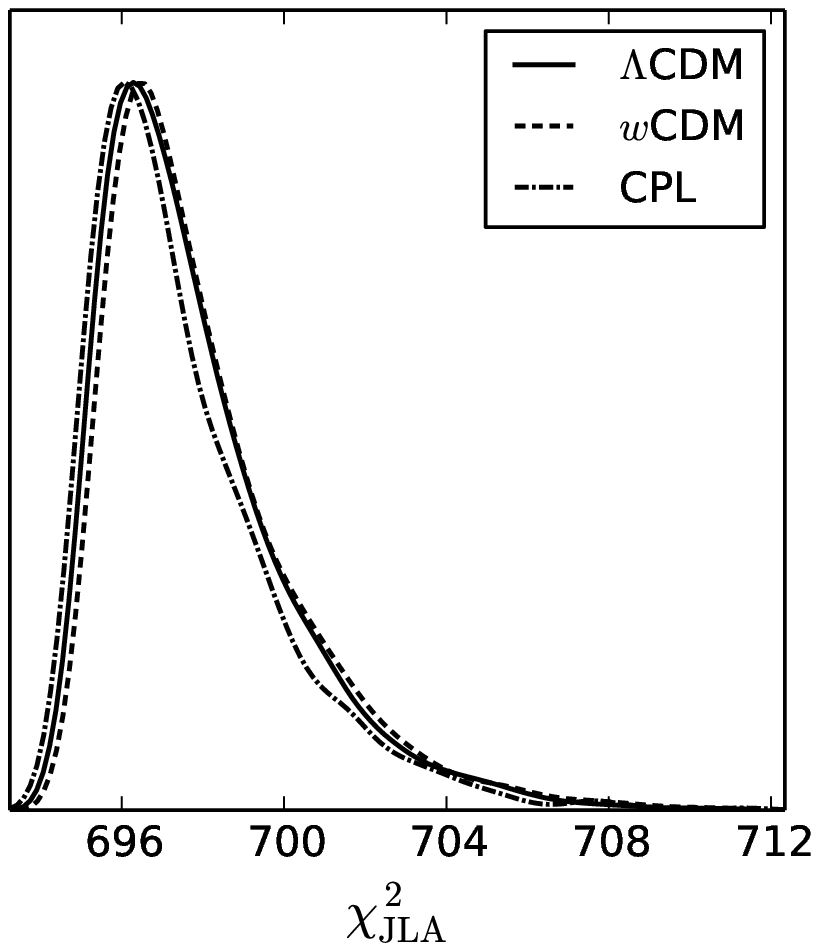}}
  \caption{Likelihood distributions for the amplitude $A_D$ and direction $(l,b)$ of dipole modulation in three cosmological models. The distribution of $\chi^2_{\rm JLA}$ is also showed in the last panel.}
  \label{fig:figure1}
\end{figure}

As was mentioned above, we study the anisotropic signals of the dipole-modulated $\Lambda$CDM, $w$CDM and CPL models by using the JLA sample. The nuisance parameters such as $\alpha$ and $\beta$ can be marginalized, since they are not model parameters with significant meanings. We just focus on studying the anisotropic signals, thus neglect the topic of model comparison. Our final results are listed in Table \ref{tab:parameters1}, where the models and the anisotropic amplitudes and preferred directions are given. The likelihood distributions of the anisotropic parameters $A_D$, $l$ and $b$ in three cosmological models are plotted in Figure~\ref{fig:figure1}. In the last panel of Figure~\ref{fig:figure1}, we also plot the distribution of $\chi^2_{\rm JLA}$.

\begin{table}
\centering
\begin{tabular}{cccc}
  \hline\hline
  parameters & $\Lambda$CDM & $w$CDM & CPL \\
  \hline
  $A_D$ & $<1.98\times10^{-3}$  & $<2.09\times10^{-3}$  & $<2.05\times10^{-3}$ \\
  $l[^{\circ}]$ & $316_{-110}^{+107}$  & $320_{-104}^{+107}$  & $318_{-183}^{+177}$ \\
  $b[^{\circ}]$ & $-5_{-60}^{+41}$  & $-4_{-61}^{+45}$  & $-8_{-54}^{+36}$ \\
  \hline
\end{tabular}
\caption{The 95\% upper bound of dipole amplitude $A_D$, and the preferred direction $(l,b)$ with $1\sigma$ uncertainty in three cosmological models.} \label{tab:parameters1}
\end{table}

In the modulated $\Lambda$CDM model, the isotropic parameter can be well constrained by using the JLA dataset only. The result is $\Omega_M=0.295\pm 0.034$ \citep{Betoule:2014frx}, which is computed for a fixed fiducial value of $H_0 = 70~{\rm km}~{\rm s}^{-1}~{\rm Mpc}^{-1}$. The change of $H_0$ does not affect the best-fit value of $\Omega_M$. By using the MCMC approach, and fixing $\Omega_M$ at 0.295, the anisotropic amplitude is constrained as $A_D<1.98\times10^{-3}$ at $95\%$ C.L., which is consistent with the isotropy within $1\sigma$ uncertainty. This implies that the JLA compilation shows no significant evidence for the deviations from isotropy. In the galactic coordinates, the dipole direction points towards F $(l,b)=(316^{\circ}\,_{-110^{\circ}}^{+107^{\circ}}, -5^{\circ}\,_{-60^{\circ}}^{+41^{\circ}})$ at $68\%$ C.L. By contrast, the Union2 sample gives constraints on the dark energy dipole as $A_D=(1.3\pm0.6)\times10^{-3}$ and $(l,b)=(309.4^\circ\pm 18.0^\circ,-15.1^\circ\pm11.5^\circ)$ at $68\%$ C.L., which implies that the anisotropic expansion of the Universe is permitted at more than $2\sigma$ \citep{Mariano:2012wx}. However, the statistical significance can be highly lowered if the full covariance matrix of SNe Ia is considered in the Union2 compilation \citep{Jimenez:2014jma}. Our constraints on the anisotropic parameters are listed in the second column in Table~\ref{tab:parameters1}. Their likelihood distributions are illustrated by the solid curves in Figure~\ref{fig:figure1}.

In the modulated $w$CDM model, there is an extra background parameter, i.e., the equation of state of dark energy $w$. This parameter can't be well constrained by using the JLA data only. The combined constraint from Planck2013+WP+BAO+JLA gives $\Omega_M=0.303\pm 0.012$, $w=-1.027\pm 0.055$ and $H_0=68.50\pm 1.27~{\rm km}~{\rm s}^{-1}~{\rm Mpc}^{-1}$ \citep{Betoule:2014frx}. By fixing these parameters at their best-fitting central values, we use the MCMC approach to constrain the anisotropic parameters. Similarly to the above discussions, we obtain the $95\%$ upper bound on anisotropic amplitude $A_D<2.09\times10^{-3}$, and preferred direction $(l,b)=(320^{\circ}\,_{-104^{\circ}}^{+107^{\circ}}, -4^{\circ}\,_{-61^{\circ}}^{+45^{\circ}})$ at $68\%$ C.L., which are listed in the third column in Table~\ref{tab:parameters1}. Their likelihood distributions are illustrated by the dashed curves in Figure~\ref{fig:figure1}.

In the modulated CPL parametrization, there are two extra background parameters, i.e., the equation of state of dark energy parameters $w_0$ and $w_1$. These two parameters also can't be well constrained by using the JLA data only. The combined constraint from Planck2013+WP+BAO+JLA gives $\Omega_M=0.304\pm 0.012$, $w_0=-0.957\pm 0.124$, $w_1=-0.336\pm 0.552$ and $H_0=68.59\pm 1.27~{\rm km}~{\rm s}^{-1}~{\rm Mpc}^{-1}$ \citep{Betoule:2014frx}. By fixing these parameters at their best-fitting central values, the anisotropic amplitude is constrained to be $A_D<2.05\times10^{-3}$ at $95\%$ C.L., and the preferred direction $(l,b)=(318^{\circ}\,_{-183^{\circ}}^{+177^{\circ}}, -8^{\circ}\,_{-54^{\circ}}^{+36^{\circ}})$ at $68\%$ C.L. These results are listed in the last column in Table~\ref{tab:parameters1}. Their likelihood distributions are illustrated by the dash-dotted curves in Figure~\ref{fig:figure1}.

From Table~\ref{tab:parameters1}, we can see that all the three cosmological models give consistent results. At first glimpse, the dipole directions we obtained seem to be consistent with that of \citet{Mariano:2012wx}. However, this is in fact not the case. In our calculation, we constrain dipole amplitude to be non-negative, i.e., $A_D\geq 0$, and let the dipole direction runs over the whole sky. \citet{Mariano:2012wx} parameterized the dipole modulation as $\mu_{\rm th}=\bar{\mu}_{\rm th}(1-A_D (\hat{\textbf{n}}\cdot\hat{\textbf{p}}))$, which has a sign difference from our parametrization of equation (\ref{muth}). Therefore, our dipole direction is actually opposite to that obtained by \citet{Mariano:2012wx}.

To test if the discrepancy between our results and previous works is due to the different dataset or different method. We apply our method to Union2 dataset, such that direct comparison with previous results can be made. The constraint on isotropic $\Lambda$CDM model from Union2 dateset gives $\Omega_M=0.274\pm 0.040$, with fiducial parameter $H_0 = 70~{\rm km}~{\rm s}^{-1}~{\rm Mpc}^{-1}$. Then we use the MCMC approach to constrain the anisotropic parameters. We obtain $A_D=(0.54_{-0.54}^{+0.13})\times 10^{-3}$ and $(l,b)=(142^{\circ}\,_{-72^{\circ}}^{+41^{\circ}}, 11^{\circ}\,_{-27^{\circ}}^{+42^{\circ}})$ at $68\%$ C.L. Regardless of the much larger uncertainty, the dipole direction we obtained is consistent with that of \citet{Mariano:2012wx}. However, our results show that the Union2 dataset is consistent with isotropy at $1\sigma$ C.L., and the $95\%$ upper bound on anisotropy is $A_D<1.40\times 10^{-3}$. On the other hand, if we apply the least-square method, we obtain $A_D=1.10\times 10^{-3}$, and $(l,b)=(126^{\circ},18^{\circ})$. Both the anisotropic amplitude and preferred direction are well consistent with the results of \citet{Mariano:2012wx}.

\section{Conclusions}\label{sec:conclusion}

In this paper, we probed the possibly anisotropic expansion of the Universe by using the recently released JLA compilation of SNe Ia. We considered the dipole-modulated deviation from the isotropy in three different dark energy models. We obtained similar constraints on the anisotropic amplitude and direction in three cases. This indicates that the preferred direction  anisotropy is insensitive to the isotropic background models. Especially, our MCMC studies show that the anisotropic amplitude has an upper bound $D<1.98\times 10^{-3}$ at $95\%$ C.L., and the dipole direction points towards $(l,b)=(316^{\circ}\,_{-110^{\circ}}^{+107^{\circ}},-5^{\circ}\,_{-60^{\circ}}^{+41^{\circ}})$ for the dipole-modulated $\Lambda$CDM model. These results imply that the there is no significant evidence for anisotropy in the JLA dataset. For comparison, we also applied MCMC method to the Union2 dataset, and we got $A_D<1.40\times 10^{-3}$ at $95\%$ C.L. The dipole direction of the Union2 points towards $(l,b)=(142^{\circ}\,_{-72^{\circ}}^{+41^{\circ}}, 11^{\circ}\,_{-27^{\circ}}^{+42^{\circ}})$ at $68\%$ C.L., which is consistent with previous results. We surprisingly found that the dipole direction derived from the JLA is approximately opposite to that from the Union2.

\section*{Acknowledgements}
We are grateful to Prof. Perivolaropoulos L. for useful comments and suggestions. This work has been funded by the National Natural Science Fund of China under grants Nos. 11375203, 11305181, 11322545, 11335012 and 11575271.

\label{lastpage}


\begin{thebibliography}{}


\bibitem[\protect\citeauthoryear{Ade et al.}{2014}]{Ade:2013nlj}Ade P. A.~R., et~al., 2014, Astron. Astrophys., 571, A23

\bibitem[\protect\citeauthoryear{Ade et al.}{2015a}]{Ade:2015xua}Ade P. A.~R., et~al., 2015a, arXiv:1502.01589

\bibitem[\protect\citeauthoryear{Ade et al.}{2015b}]{Ade:2015lrj}Ade P. A.~R., et~al., 2015b, arXiv:1502.02114

\bibitem[\protect\citeauthoryear{Akrami et al.}{2014}]{Akrami:2014eta}Akrami Y., Fantaye Y., Shafieloo A., Eriksen H. K., Hansen F. K., Banday A. J., G\'{o}rski K. M., 2014, Astrophys.\ J., 784, L42

\bibitem[\protect\citeauthoryear{Amanullah et al.}{2010}]{Amanullah:2010vv}Amanullah R., Lidman C., Rubin D., Aldering G., Astier P., et~al., 2010, Astrophys. J., 716, 712


\bibitem[\protect\citeauthoryear{Antoniou \& Perivolaropoulos}{2010}]{Antoniou:2010gw}Antoniou I., Perivolaropoulos L., 2010, JCAP, 1012, 012


\bibitem[\protect\citeauthoryear{Armendariz-Picon}{2004}]{ArmendarizPicon:2004pm}Armendariz-Picon C., 2004, JCAP, 0407, 007


\bibitem[\protect\citeauthoryear{Bengaly, Bernui \& Alcaniz}{2015}]{Bengaly:2015dza}Bengaly C. A.~P., Bernui A., Alcaniz J.~S., 2015, Astrophys. J., 808, 39

\bibitem[\protect\citeauthoryear{Bennett et al.}{2013}]{Bennett:2012zja}Bennett C.~L., et~al., 2013, Astrophys. J., 208, S20

\bibitem[\protect\citeauthoryear{Betoule et al.}{2014}]{Betoule:2014frx}Betoule M., et~al., 2014, Astron. Astrophys., 568, A22

\bibitem[\protect\citeauthoryear{Bielewicz, Gorski \& Banday}{2004}]{Bielewicz:2004en}Bielewicz P., Gorski K.~M., Banday A.~J., 2004, Mon. Not. Roy. Astron. Soc., 355, 1283

\bibitem[\protect\citeauthoryear{Blomqvist, Enander \& Mortsell}{2010}]{Blomqvist:2010ky}Blomqvist M., Enander J., Mortsell E., 2010, JCAP, 1010, 018

\bibitem[\protect\citeauthoryear{Cai et al.}{2013}]{Cai:2013lja}Cai R.-G., Ma Y.-Z., Tang B., Tuo Z.-L., 2013, Phys. Rev. D, 87, 123522

\bibitem[\protect\citeauthoryear{Cai \& Tuo}{2012}]{Cai:2011xs}Cai R.-G., Tuo Z.-L., 2012, JCAP, 1202, 004

\bibitem[\protect\citeauthoryear{Campanelli et al.}{2011}]{Campanelli:2010zx}Campanelli L., Cea P., Fogli G.~L., Marrone A., 2011, Phys. Rev. D, 83, 103503

\bibitem[\protect\citeauthoryear{Chang et al.}{2013}]{Chang:2013xwa}Chang Z., Li M.-H., Li X., Wang S., 2013, Eur. Phys. J. C, 73, 2459

\bibitem[\protect\citeauthoryear{Chang, Li \& Wang}{2013}]{Chang:2013zwa}Chang Z., Li M.-H., Wang S., 2013, Phys. Lett. B, 723, 257

\bibitem[\protect\citeauthoryear{Chang et al.}{2014a}]{Chang:2014wpa}Chang Z., Li X., Lin H.-N., Wang S., 2014a, Eur. Phys. J. C, 74, 2821

\bibitem[\protect\citeauthoryear{Chang et al.}{2014b}]{Chang:2014jza}Chang Z., Li X., Lin H.-N., Wang S., 2014b, Mod. Phys. Lett. A, 29, 1450067

\bibitem[\protect\citeauthoryear{Chang \& Lin}{2015}]{Chang:2014nca}Chang Z., Lin H.-N., 2015, Mon. Not. Roy. Astron. Soc., 446, 2952

\bibitem[\protect\citeauthoryear{Chevallier \& Polarski}{2001}]{Chevallier:2000qy}Chevallier M., Polarski D., 2001, Int. J. Mod. Phys. D, 10, 213

\bibitem[\protect\citeauthoryear{Colin et al.}{2011}]{Colin:2010ds}Colin J., Mohayaee R., Sarkar S., Shafieloo A., 2011, Mon. Not. Roy. Astron. Soc., 414, 264

\bibitem[\protect\citeauthoryear{Copi et~al.}{2010}]{Copi:2010na}Copi C.~J., Huterer D., Schwarz D.~J., Starkman G.~D., 2010, Adv. Astron., 2010, 847541

\bibitem[\protect\citeauthoryear{Dodelson}{2003}]{Dodelson:2003ft}Dodelson S., 2003, Modern cosmology. Amsterdam, Netherlands: Academic Pr.

\bibitem[\protect\citeauthoryear{Eriksen et al.}{2004}]{Eriksen:2003db}Eriksen H. K., Hansen F. K., Banday A. J., Gorski K. M., Lilje P. B., 2004, Astrophys.\ J., 605, 14 [Erratum-ibid., 609, 1198]

\bibitem[\protect\citeauthoryear{Frommert \& En{\ss}lin}{2010}]{Frommert:2009qw}Frommert M., En{\ss}lin T.~A., 2010, Mon. Not. Roy. Astron. Soc., 403, 1739

\bibitem[\protect\citeauthoryear{Gupta, Saini \& Laskar}{2008}]{Gupta:2007pb}Gupta S., Saini T.~D., Laskar T., 2008, Mon. Not. Roy. Astron. Soc., 388, 242

\bibitem[\protect\citeauthoryear{Hansen, Banday \& Gorski}{2004}]{Hansen:2004vq}Hansen F. K., Banday A. J., Gorski K. M., 2004, Mon.\ Not.\ Roy.\ Astron.\ Soc., 354, 641

\bibitem[\protect\citeauthoryear{Heneka, Marra \& Amendola}{2014}]{Heneka:2013hka}Heneka C., Marra V., Amendola L., 2014, Mon.\ Not.\ Roy.\ Astron.\ Soc., 439, 1855

\bibitem[\protect\citeauthoryear{Hinshaw et al.}{2013}]{Hinshaw:2012aka}Hinshaw G.,  et~al., 2013, Astrophys. J., 208, S19

\bibitem[\protect\citeauthoryear{Hutsemekers et~al.}{2005}]{Hutsemekers:2005iz}Hutsemekers D., Cabanac R., Lamy H., Sluse D., 2005, Astron. Astrophys., 441, 915

\bibitem[\protect\citeauthoryear{Hutsemekers \& Lamy}{2001}]{Hutsemekers:2000fv}Hutsemekers D., Lamy H., 2001, Astron. Astrophys., 367, 381

\bibitem[\protect\citeauthoryear{Jimenez, Salzano \& Lazkoz}{2015}]{Jimenez:2014jma}Jimenez J.~B., Salzano V., Lazkoz R., 2015, Phys. Lett. B, 741, 168

\bibitem[\protect\citeauthoryear{Kalus et al.}{2013}]{Kalus:2012zu}Kalus B., Schwarz D.~J., Seikel M., Wiegand A., 2013, Astron. Astrophys., 553, A56

\bibitem[\protect\citeauthoryear{Kashlinsky et al.}{2010}]{Kashlinsky:2009dw}Kashlinsky A., Atrio-Barandela F., Ebeling H., Edge A., Kocevski D., 2010, Astrophys. J., 712, L81

\bibitem[\protect\citeauthoryear{Kashlinsky et al.}{2009}]{Kashlinsky:2008ut}Kashlinsky A., Atrio-Barandela F., Kocevski D., Ebeling H., 2009, Astrophys. J., 686, L49

\bibitem[\protect\citeauthoryear{King et~al.}{2012}]{King:2012id}King J.~A., Webb J.~K., Murphy M.~T., Flambaum V.~V., et~al., 2012, Mon. Not. Roy. Astron. Soc., 422, 3370

\bibitem[\protect\citeauthoryear{Koivisto \& Mota}{2008}]{Koivisto:2008xf}Koivisto T. S., Mota D. F., 2008, JCAP, 0808, 021

\bibitem[\protect\citeauthoryear{Koivisto et al.}{2011}]{Koivisto:2010dr}Koivisto T. S., Mota D. F., Quartin M., Zlosnik T. G., 2011, Phys.\ Rev.\ D, 83, 023509

\bibitem[\protect\citeauthoryear{Lewis \& Bridle}{2002}]{Lewis:2002ah}Lewis A., Bridle S., 2002, Phys. Rev. D, 66, 103511

\bibitem[\protect\citeauthoryear{Li et al.}{2013}]{Li:2013vea}Li X., Lin H.-N., Wang S., Chang Z., 2013, Eur. Phys. J. C, 73, 2653

\bibitem[\protect\citeauthoryear{Li et al.}{2015}]{Li:2015uda}Li X., Lin H.-N., Wang S., Chang Z., 2015, Eur. Phys. J. C, 75, 181

\bibitem[\protect\citeauthoryear{Linder}{2003}]{Linder:2002et}Linder E.~V., 2003, Phys. Rev. Lett., 90, 091301

\bibitem[\protect\citeauthoryear{Mariano \& Perivolaropoulos}{2012}]{Mariano:2012wx}Mariano A., Perivolaropoulos L., 2012, Phys. Rev. D, 86, 083517

\bibitem[\protect\citeauthoryear{Mariano \& Perivolaropoulos}{2013}]{Mariano:2012ia}Mariano A., Perivolaropoulos L., 2013, Phys. Rev. D, 87, 043511

\bibitem[\protect\citeauthoryear{Molaro et al.}{2013}]{Molaro:2013saa}Molaro P., Centurion M., Whitmore J.~B., Evans T. M., Murphy M. T., et al., 2013, Astron. Astrophys., 555, A68

\bibitem[\protect\citeauthoryear{Perivolaropoulos}{2008}]{Perivolaropoulos:2008ud}Perivolaropoulos L., 2008, arXiv:0811.4684

\bibitem[\protect\citeauthoryear{Perivolaropoulos}{2011}]{Perivolaropoulos:2011hp}Perivolaropoulos L., 2011, J. Cosmol., 15, 6054

\bibitem[\protect\citeauthoryear{Quartin \& Notari}{2015}]{Quartin:2014yaa}Quartin M., Notari A., 2015, JCAP, 1501, 008


\bibitem[\protect\citeauthoryear{Salehi \& Aftabi}{2015}]{Salehi:2015ira}Salehi A., Aftabi S., 2015, arXiv:1502.04507

\bibitem[\protect\citeauthoryear{Schucker, Tilquin \& Valent}{2014}]{Schucker:2014wca}Schucker T., Tilquin A., Valent G., 2014, Mon. Not. Roy. Astron. Soc., 444, 2820

\bibitem[\protect\citeauthoryear{Schwarz \& Weinhorst}{2007}]{Schwarz:2007wf}Schwarz D.~J., Weinhorst B.,  2007, Astron. Astrophys., 474, 717

\bibitem[\protect\citeauthoryear{Suzuki et~al.}{2012}]{Suzuki:2011hu}Suzuki N., Rubin D., Lidman C., Aldering G., Amanullah R., et~al., 2012, Astrophys. J., 746, 85

\bibitem[\protect\citeauthoryear{Tegmark, de Oliveira-Costa \& Hamilton}{2003}]{Tegmark:2003ve}Tegmark M., de Oliveira-Costa A., Hamilton A., 2003, Phys. Rev. D, 68, 123523

\bibitem[\protect\citeauthoryear{Wang \& Wang}{2014}]{Wang:2014vqa}Wang J.~S., Wang F.~Y., 2014, Mon. Not. Roy. Astron. Soc., 443, 1680

\bibitem[\protect\citeauthoryear{Watkins, Feldman \& Hudson}{2009}]{Watkins:2008hf}Watkins R., Feldman H.~A., Hudson M. J., 2009, Mon. Not. Roy. Astron. Soc., 392, 743

\bibitem[\protect\citeauthoryear{Webb et~al.}{2011}]{Webb:2010hc}Webb J. K., King J. A., Murphy M. T., Flambaum V. V., Carswell R. F., et al., 2011, Phys. Rev. Lett., 107, 191101

\bibitem[\protect\citeauthoryear{Yang, Wang \& Chu}{2014}]{Yang:2013gea}Yang X., Wang F. Y., Chu Z., 2014, Mon. Not. Roy. Astron. Soc., 437, 1840

\bibitem[\protect\citeauthoryear{Zhao, Wu \& Zhang}{2013}]{Zhao:2013yaa}Zhao W., Wu P.~X., Zhang Y., 2013, Int. J. Mod. Phys. D, 22, 1350060

\end{thebibliography}
\end{document}